\documentclass[a4paper, conference]{IEEEtran}
\usepackage[T1]{fontenc}
\IEEEoverridecommandlockouts
\usepackage{cite}
\usepackage{amsmath,amssymb,amsfonts}
\usepackage{algorithmic}
\usepackage{graphicx}
\usepackage{textcomp}
\usepackage{xcolor}
\usepackage{mathtools, stackengine, mathrsfs, setspace, bm, multirow, adjustbox, calc, tikz, pgfplots, environ, xcolor, multirow}
\PassOptionsToPackage{hyphens}{url}
\usepackage[bookmarks=false]{hyperref}
\usepgfplotslibrary{fillbetween}
\usepackage{xparse}
\pgfplotsset{width=8.1cm, legend style={font=\small}}
\newsavebox{\fminipagebox}
\NewDocumentEnvironment{fminipage}{m O{\fboxsep}}
 {\par\kern#2\noindent\begin{lrbox}{\fminipagebox}
  \begin{minipage}{#1}\ignorespaces}
 {\end{minipage}\end{lrbox}%
  \makebox[#1]{%
    \kern\dimexpr-\fboxsep-\fboxrule\relax
    \fbox{\usebox{\fminipagebox}}%
    \kern\dimexpr-\fboxsep-\fboxrule\relax
  }\par\kern#2
 }
\begingroup\edef\x{\endgroup
  \mathchardef\mathdollar=\the\numexpr"7000+\the\mathdollar\relax
}\x
\DeclareMathAlphabet{\mathit}{T1}{cmr}{m}{it}

\makeatletter
\let\MYcaption\@makecaption
\makeatother
\usepackage{caption}
\DeclareCaptionLabelSeparator{colonquad}{:\quad}
\DeclareCaptionStyle{fig-top}%
	[justification=raggedright,indention=2cm, font=footnotesize, labelsep=colonquad]{}
\DeclareCaptionStyle{fig}%
	[justification=raggedright,indention=2cm, font=footnotesize, labelsep=colonquad]{}	
\makeatletter
\let\@makecaption\MYcaption
\makeatother

\def\BibTeX{{\rm B\kern-.05em{\sc i\kern-.025em b}\kern-.08em
    T\kern-.1667em\lower.7ex\hbox{E}\kern-.125emX}}

\begin{document}
\bstctlcite{IEEEexample:BSTcontrol}
\title{Forecasting Daily Primary Three-Hour \\Net Load Ramps in the \textit{CAISO} System
\thanks{This work was supported in part by the Research Council of Norway under the ``LUCS'' project, and by the German Federal Ministry for Economic Affairs and Energy under Grant 03EI6004B.}
}
\IEEEoverridecommandlockouts
\author{\IEEEauthorblockN{Ogun Yurdakul,
Andreas Meyer,
Fikret Sivrikaya, and
Sahin Albayrak}
\IEEEauthorblockA{Department of Electrical Engineering and Computer Science\\
Technical University of Berlin, Berlin, Germany}
Email: \{yurdakul, andreas.meyer, fikret.sivrikaya, sahin.albayrak\}@tu-berlin.de\\
\small{\textit{accepted for presentation in the 52$^{\text{nd}}$ North American Power Symposium}}
\vspace{-0.5cm}}
\maketitle
\IEEEpubidadjcol
\begin{abstract}
The deepening penetration of variable energy resources creates unprecedented challenges for system operators (\textit{SO}s). An issue that merits special attention is the precipitous net load ramps, which require \textit{SO}s to have flexible capacity at their disposal so as to maintain the supply-demand balance at all times. In the judicious procurement and deployment of flexible capacity, a tool that forecasts net load ramps may be of great assistance to \textit{SO}s. To this end, we propose a methodology to forecast the magnitude and start time of daily primary three-hour net load ramps. We perform an extensive analysis so as to identify the factors that influence net load and draw on the identified factors to develop a forecasting methodology that harnesses the long short-term memory model. We demonstrate the effectiveness of the proposed methodology on the \textit{CAISO} system using comparative assessments with selected benchmarks based on various evaluation metrics.
\end{abstract}
\section{Introduction}\label{1}
The deepening penetration of variable energy resources (\textit{VER}s) has given rise to the use of the legacy grid in a way different from that for which it was designed, thereby posing unprecedented challenges for system operators (\textit{SO}s). One such challenge is the rapid diurnal fluctuation of net load, that is, system load less photovoltaic (\textit{PV}) generation less wind generation.\footnote{The \textit{PV} generation and wind generation referred to in the net load definition pertain to those from the utility-scale installations, as those from behind-the-meter installations are taken into account in system load.} \par
\begin{figure}[h]
\centering
\vspace{-0.2cm}
\includegraphics[keepaspectratio, height=4.0cm]{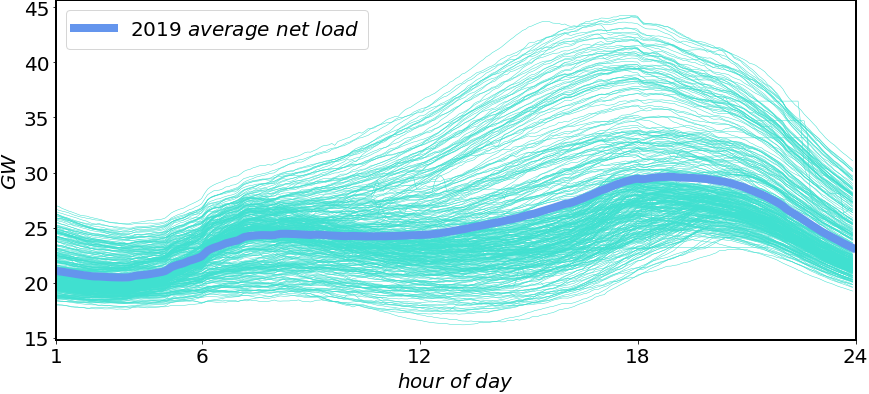}
\vspace{-0.5cm}
\caption{\textit{CAISO} system daily net load in 2019}
\vspace{-0.2cm}
\label{fig1}
\end{figure}

\begin{table*}
\begin{minipage}[c][8.5cm][t]{\linewidth}
\begin{fminipage}{\textwidth}[1ex]
\footnotesize
\renewcommand{\arraystretch}{1.2}
\begin{tabular}{l l l l}
\multicolumn{2}{l}{\textbf{Nomenclature}}   & 	 &\\
$\mathbb{R}^{n}$ & set of real-valued $n-$dimensional column vectors \hspace{0.85cm} &	$\boldsymbol{\tilde{V}}^{\mathsf{\alpha}}_{\delta} \in \mathbb{R}^{\mathcal{T}\times4}$ &matrix of values obtained after the application of  \\
$\mathbb{R}^{m \times n}$ & set of $m-$by$-n$ real-valued matrices &	& min-max scaling to each corresponding element of $\boldsymbol{V}^{\mathsf{\alpha}}_{\delta}$ \\
$\mathcal{T}$ & length of input sequence, \textit{look-back} window &	$\boldsymbol{V}^{\mathsf{\gamma}}_{\delta}\in\mathbb{R}^{\mathcal{T}\times5}$ &matrix of historical solar irradiance measurements from\\
$\delta$ &  day for which three-hour net load ramp is forecast &	& the selected five weather stations for $\mathcal{T}$ time periods\\
$\boldsymbol{p}^{\mathsf{n}}_{\delta}\in\mathbb{R}^{\mathcal{T}}$ &vector of historical net load values &	$\boldsymbol{\tilde{V}}^{\mathsf{\gamma}}_{\delta} \in \mathbb{R}^{\mathcal{T}\times5}$ & matrix of values obtained after the application of  \\
& for $\mathcal{T}$ time periods &	& min-max scaling to each corresponding element of $\boldsymbol{V}^{\mathsf{\gamma}}_{\delta}$ \\
$\boldsymbol{\tilde{p}}^{\mathsf{n}}_{\delta} \in \mathbb{R}^{\mathcal{T}}$ & vector of values obtained after the application of  &	$\boldsymbol{d}_{\delta}\in\mathbb{R}^{\mathcal{T}}$ &vector of day of week of the $\mathcal{T}$ time periods \\
& min-max scaling to each corresponding element of $\boldsymbol{p}^{\mathsf{n}}_{\delta}$ &	$\boldsymbol{\tilde{d}}_{\delta} \in \mathbb{R}^{\mathcal{T}}$ & one-hot encoding representation of $\boldsymbol{d}_{\delta}$ \\
$\boldsymbol{{p}}^{\mathsf{\ell}}_{\delta} \in \mathbb{R}^{\mathcal{T}}$ & vector of historical power system load measurements &	$\boldsymbol{m}_{\delta}\in\mathbb{R}^{\mathcal{T}}$ &vector of the month of year of the $\mathcal{T}$ time periods  \\
& for $\mathcal{T}$ time periods &	$\boldsymbol{\tilde{m}}_{\delta} \in \mathbb{R}^{\mathcal{T}}$ & one-hot encoding representation of $\boldsymbol{m}_{\delta}$\\
$\boldsymbol{\tilde{p}}^{\mathsf{\ell}}_{\delta} \in \mathbb{R}^{\mathcal{T}}$ & vector of values obtained after the application of  &	$\hat{\mu}_{\delta}$ & forecast magnitude of the primary three-hour net load  \\
& min-max scaling to each corresponding element of $\boldsymbol{p}^{\mathsf{\ell}}_{\delta}$ &	&ramp on day $\delta$ \\
$\boldsymbol{p}^{\mathsf{s}}_{\delta}\in\mathbb{R}^{\mathcal{T}}$ &vector of historical \textit{PV} generation measurements&	$\mu_{\delta}$  & actual magnitude of the primary three-hour net \\
& for $\mathcal{T}$ time periods &	& load ramp on day $\delta$ \\
$\boldsymbol{\tilde{p}}^{\mathsf{s}}_{\delta} \in \mathbb{R}^{\mathcal{T}}$ & vector of values obtained after the application of  &	$\hat{\kappa}_{\delta}$ & forecast time period at which the primary\\
& min-max scaling to each corresponding element of $\boldsymbol{p}^{\mathsf{s}}_{\delta}$ &	& three-hour net load ramp on day $\delta$ starts \\
$\boldsymbol{p}^{\mathsf{w}}_{\delta}\in\mathbb{R}^{\mathcal{T}}$ &vector of historical wind generation measurements &	$\kappa_{\delta}$ & actual time period at which the primary\\
& for $\mathcal{T}$ time periods &	& three-hour net load ramp on day $\delta$ starts \\
$\boldsymbol{\tilde{p}}^{\mathsf{w}}_{\delta} \in \mathbb{R}^{\mathcal{T}}$ & vector of values obtained after the application of  &	$\boldsymbol{X}_{\delta}$ & input matrix\\
& min-max scaling to each corresponding element of $\boldsymbol{p}^{\mathsf{w}}_{\delta}$ &	$(\cdot)^{\mathsf{T}}$ & transposition \\
$\boldsymbol{V}^{\mathsf{\alpha}}_{\delta}\in\mathbb{R}^{\mathcal{T}\times4}$ &matrix of historical temperature measurements from&	$\mathcal{N}$ & number of training samples \\
& the selected four weather stations for $\mathcal{T}$ time periods&	&
\end{tabular}
\end{fminipage}
\end{minipage}
\end{table*}
In California, the net load has a morning peak, falls off around noon, and rapidly picks up in the late afternoon, creating a graph that resembles the silhouette of a duck---hence the sobriquet “duck curve” coined by the California Independent System Operator (\textit{CAISO}). Specifically, the net load attains markedly low values around midday due to relatively low demand and high \textit{PV} generation, and it rises rapidly in the afternoon as the load picks up and the \textit{PV} generation decreases. We present in Fig. \ref{fig1} the daily net load plots in the \textit{CAISO} system in 2019, on which two distinct ramps can be identified: one ramp of mostly larger magnitude in the afternoon, and the other ramp generally occurring in the morning. For any given day, the maximum change in net load over a three-hour time window is referred to as the primary three-hour net load ramp. The magnitude of daily primary three-hour net load ramps may hit upwards of 15,000 \textit{MW} and reach 50\% of the daily peak load \cite{rampbib:10}.\par
Net load ramps of such markedly high magnitudes warrant and call for explicit flexible capacity requirements to ensure that supply-demand balance be maintained around the clock. For this purpose, the \textit{CAISO} undertakes the so-called \textit{flexible capacity needs assessment}, which takes into account the load, \textit{PV} generation, and wind generation measurements in the previous year---in conjunction with the additional \textit{PV} and wind installations and load growth factor for the current year---to generate net load estimates for the subsequent year \cite{rampbib:8}. The net load estimates are jointly evaluated with the contingency reserve obligations to stipulate monthly minimum flexible capacity requirements. Based on the estimated temporal distribution of net load ramps, the minimum flexible capacity requirements are set forth under three categories, \textit{viz.}: base flexibility, peak flexibility, and super-peak flexibility, among which peak flexibility and super-peak flexibility categories may entail different must-offer hours for each month \cite{rampbib:7}. \par
The uncertainty in system load, compounded by the uncertain, highly time-varying, and intermittent nature of \textit{VER}s, renders the performance of net load forecasts a formidable challenge \cite{rampbib:11}. Indeed, the stipulated average flexible capacity requirements fell short of the magnitude of the actual primary three-hour net load ramps in the eight months of 2018 \cite{rampbib:8}. The net load plots in Fig. \ref{fig1} further make evident that \textit{one-size-fits-all} monthly requirements cannot sufficiently capture the idiosyncrasies of the magnitude or start time of daily primary three-hour net load ramps. \par
The \textit{CAISO} makes use of other mechanisms to insure that, if need be, sufficient flexible capacity be readily deployable to meet steep ramps, including residual unit commitment and load forecast adjustment \cite{rampbib:8}. The manual increase of hour-ahead and 15-minute load forecasts brings about higher imports, thereby affording additional flexible generation from within the \textit{CAISO} system. A market-based mechanism, designed specifically to address the uncertainty in net load, is the flexible ramping product, which procures flexible ramping capacity in the real-time market so as to obviate the need for manual load adjustments. \par
The effective implementation of flexible capacity requirements can be greatly aided by a tool that forecasts the magnitude and start time of daily primary three-hour net load ramps. In short-term operations, \textit{SO}s can capitalize on such a tool to judiciously decide on and effectuate the necessary course of actions so as to maintain the supply-demand balance at all times.\par
In this paper, we propose such a methodology to forecast the magnitude and start time of the primary three-hour net load ramp on the subsequent day. The proposed methodology is based on the long short-term memory (\textit{LSTM}) model, which is a recurrent neural network (\textit{RNN}) architecture with the hallmark capability to capture long-term temporal dependencies.\par
While the proposed methodology can be tailored for any power system, a specific application of the methodology would be remiss, if it were to disregard the key characteristics of the system to which it is applied. On the grounds that we illustrate the application of the proposed methodology on the \textit{CAISO} system, we hone in on the salient features of the net load of the \textit{CAISO} system and examine the components of net load, \textit{viz.}: system load, \textit{PV} generation, and wind generation. We rely heavily on the body of empirical observations to analyze the variation of each component over time so as to garner insights into the underlying phenomena that affect net load. These insights are explicitly leveraged in the design of the proposed methodology.
\subsection{Related Work}
There is a growing body of literature on net load forecasting. Reference \cite{rampbib:17} leverages Bayesian theory in conjunction with \textit{LSTM} networks to develop a probabilistic net load forecasting model. In \cite{rampbib:16}, a probabilistic net load forecasting methodology, based on dependent discrete convolution, is developed. Deep neural networks are harnessed in \cite{rampbib:18} to develop a net load forecasting model, whose features include historical net load data, meteorological data, and electricity prices. \par
A line of research similar to net load ramp forecasting is wind power ramp forecasting, i.e., forecasting the rapid fluctuations in wind power generation. An ensemble method comprising empirical mode decomposition, kernel ridge regression, and random vector functional link neural network is developed in \cite{rampbib:1} for wind power ramp rate forecasting. In \cite{rampbib:3}, a hybrid forecasting model to forecast wind power ramps is proposed, where the authors utilize orthogonal test jointly with support vector machine.
\subsection{Contributions and structure of the paper}
The general contributions and novel aspects of this paper are as follows:
\begin{enumerate}
\item We develop a novel \textit{LSTM}-based methodology to forecast the magnitude and start time of the primary three-hour net load ramp on the subsequent day.\footnote{The source code is available at https://github.com/oyurdakul/ramp.} We demonstrate the effectiveness of the proposed methodology on real-world \textit{CAISO} data using various benchmarks and evaluation metrics. Given the dearth of studies in this area, the present paper may serve as a jumping-off point for future studies in this field. 
\item We provide valuable insights into the influence of the temporal variation of system load, \textit{PV} generation, and wind generation, on the magnitude and start time of daily primary three-hour net load ramps. We effectively exploit the gained insights to pinpoint the relevant features of the designed \textit{LSTM} network.
\item The proposed methodology makes a dent in the development of a decision support tool that can assist \textit{SO}s in the procurement and deployment of flexible capacity to meet steep and sustained net load ramps. 
\end{enumerate}
The remainder of the paper consists of four sections. In Section \ref{2}, we investigate the factors that influence daily primary three-hour net load ramps in the \textit{CAISO} system and identify the relevant features of the proposed forecasting methodology. We briefly describe \textit{LSTM} in Section \ref{3} and present the proposed forecasting methodology. In Section \ref{4}, we illustrate the application of the proposed methodology to forecast the magnitude and start time of the primary three-hour net load ramp in the \textit{CAISO} system on the subsequent day and discuss the results. We present concluding remarks and discuss the scope of further work in Section \ref{5}. 
\section{Net Load Variation Assessment}\label{2}
In this section, we present a systematic exposition of the factors that influence net load so as to identify the pertinent features of a methodology to forecast the magnitude and start time of daily primary three-hour net load ramps. As per the application of the proposed methodology in Section \ref{4}, we focus the scope of our discussion in this section on the \textit{CAISO} system. \par
\begin{figure}[h]
\centering
\vspace{-0.2cm}
\includegraphics[width=8.89cm, height=4cm]{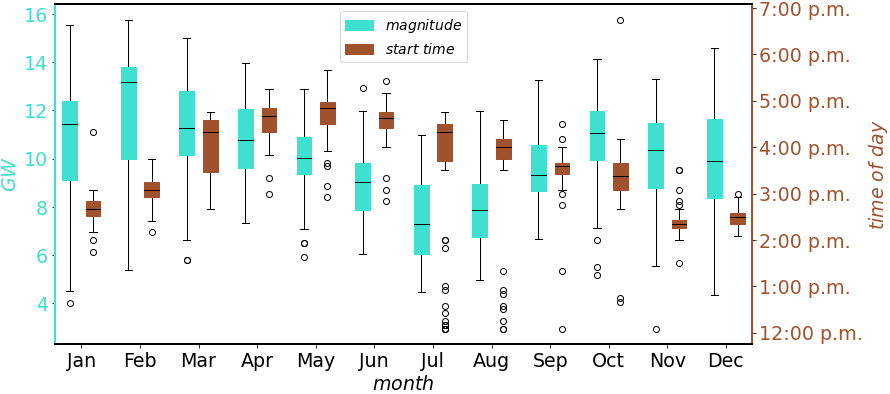}
\vspace{-0.5cm}
\caption{\textit{CAISO} system start time and magnitude of daily primary three-hour net load ramps in 2019}
\vspace{-0.3cm}
\label{fig2}
\end{figure}
Fig. \ref{fig2} presents the magnitude and start time of daily primary three-hour net load ramps in the \textit{CAISO} system in 2019 and makes evident a number of general temporal patterns. We observe from Fig. \ref{fig2} that the magnitude of daily primary three-hour net load ramps in the \textit{CAISO} system, by and large, attains lower values in summer. Further, the start time of daily primary three-hour net load ramps in the \textit{CAISO} system in the summer months is mostly later in the day than that in the non-summer months. To substantiate these empirical observations and gain the capability to capture the idiosyncrasies of daily primary three-hour net load ramps, we undertake a thorough analysis of the components of net load.\par
\begin{figure}[h]
\vspace{-0.2cm}
\includegraphics[width=8.89cm, height=4cm]{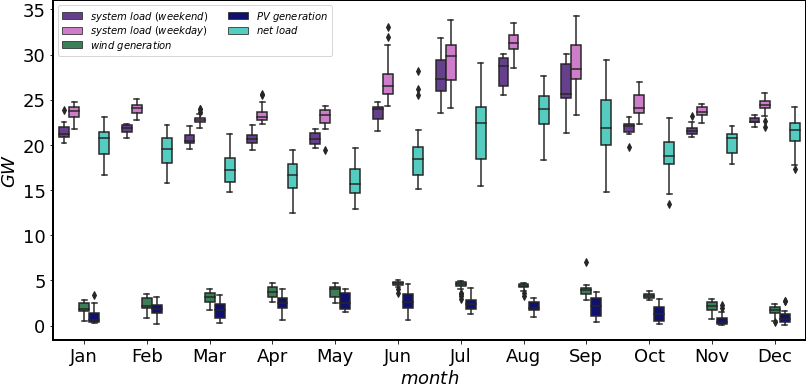}
\vspace{-0.5cm}
\caption{\textit{CAISO} average net load, system load, \textit{PV} generation, and wind generation in 2019}
\vspace{-0.3cm}
\label{fig3}
\end{figure}
The box-whisker plots of \textit{CAISO} system load provided in Fig. \ref{fig3} reflect that the \textit{CAISO} system load attains higher values in summer months and on weekdays in comparison with the non-summer months and weekends, respectively. A key determinant of system load, brought on especially due to the ubiquity of heating, ventilation, and air-conditioning systems in California, is temperature. We conduct experiments with temperature data collected from different weather stations located in the densely populated regions of California and pick four weather stations in San Francisco, San Jose, Los Angeles, and Sacramento.
\begin{figure}[h]
\centering
\includegraphics[keepaspectratio, height=3.5cm]{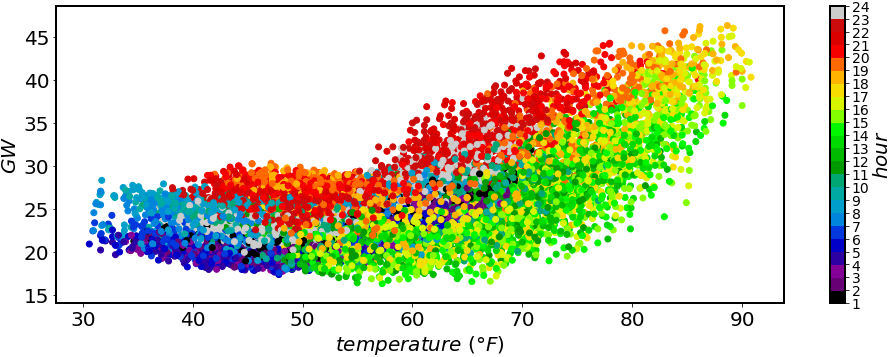}
\vspace{-0.1cm}
\caption{Correlation between temperature and system load}
\label{fig4}
\vspace{-0.1cm}
\end{figure}
Fig. \ref{fig4} bears out the correlation between the \textit{CAISO} system load and the temperature measurements collected from the said weather stations. For temperature values above around 60 degrees, the \textit{CAISO} system load rises with increasing temperature, which may be attributed to the higher power consumption of air conditioners. On the flip side, the \textit{CAISO} system load also increases---albeit at a slower rate---as the temperature falls below 60 degrees, which may be ascribed to the power consumption of electric heaters.\par
\begin{figure}[h]
\vspace{-0.2cm}
\centering
\includegraphics[height=3.3cm, keepaspectratio]{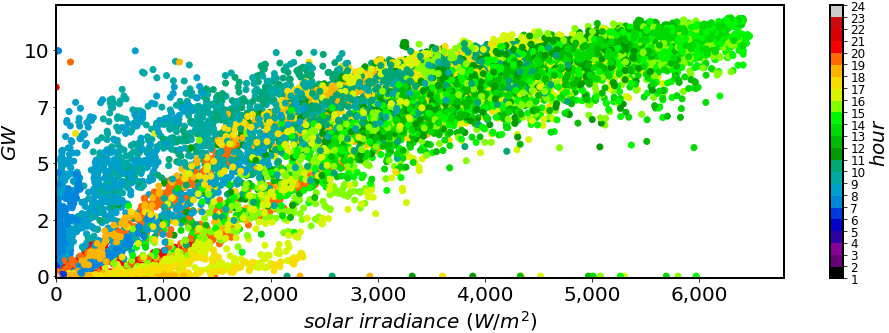}
\vspace{-0.3cm}
\caption{Correlation between solar irradiance and \textit{PV} generation}
\label{fig5}
\vspace{-0.2cm}
\end{figure}

We next turn our investigation to the \textit{PV} generation in the \textit{CAISO} system. We observe from Fig. \ref{fig3} that the \textit{CAISO} \textit{PV} generation in summer months is typically higher than that in the non-summer months. Further, in light of the major influence of solar irradiance on \textit{PV} generation, we experiment with solar irradiance measurements collected from different weather stations by assessing their location vis-à-vis the spatial distribution of \textit{PV} installations across California. We settle on five weather stations located in Rosamond, Lancaster, Desert Center, and Santa Margarita, among which Santa Margarita contains two designated weather stations so as to expressly capture the solar irradiance measurements in the Topaz Solar Farm and California Valley Solar Ranch. We depict the correlation between the the solar irradiance measurements collected from the designated weather stations and \textit{CAISO} \textit{PV} generation in Fig. \ref{fig5}, which corroborates the tight coupling between solar irradiance and \textit{PV} generation. Analogously, the diversity in \textit{CAISO} wind generation levels over time are borne out in Fig. \ref{fig3}. \par
The temporal variation of net load and its components elucidated in this section brings out the need that the proposed methodology explicitly incorporate as features the historical net load, system load, \textit{PV} generation, and wind generation measurements, solar irradiance and temperature measurements from the said weather stations, and the associated temporal information. In the next section, we spell out the specifics as to how these data/information are effectively exploited by the proposed forecasting methodology.
\section{Net Load Ramp Forecasting Methodology}\label{3}
The analysis in Section \ref{2} highlights the key factors that influence net load and punctuates the necessity of the consideration of time in the development of a methodology to forecast the magnitude and start time of daily primary three-hour net load ramps. Indeed, a major requirement of such a methodology is the ability to effectively grasp the intertemporal relationships among data points. To this end, we harness the \textit{LSTM} model, whose salient capability is to learn long-term temporal dependencies.\par
\subsection{Long Short-Term Memory Model}
An \textit{LSTM} block includes three \textit{gates}: the input gate, the forget gate, and the output gate, which control the information flow of an \textit{LSTM} block. A key architectural element of an \textit{LSTM} block is the \textit{memory cell}, which, in conjunction with the operation of the gates, spearheads the storage of valuable information. The joint operation of the memory cell and the \textit{LSTM} gates imparts the capability to capture long-term temporal dependencies to an \textit{LSTM} block {\cite{rampbib:15}}.
\subsection[title]{Proposed Methodology}
The proposed methodology harnesses two \textit{LSTM} networks---one for each of the two tasks of forecasting the magnitude and start time of the primary three-hour net load ramp on the subsequent day. The input features of the two networks are the same and include the historical net load, system load, \textit{PV} generation, and wind generation measurements, as well as the ambient temperature and solar irradiance measurements from the designated weather stations for $\mathcal{T}$ time periods, where the term $\mathcal{T}$ denotes the so-called \textit{look-back} window. To explicitly take into account the temporal information, we also provide the \textit{LSTM} networks with the month of the year and the day of the week of the selected $\mathcal{T}$ time periods. Since \textit{LSTM} networks are sensitive to the scale of the input data, we start out by preprocessing the data of the input features.\par
\begin{figure}[h]
\includegraphics[width=\linewidth]{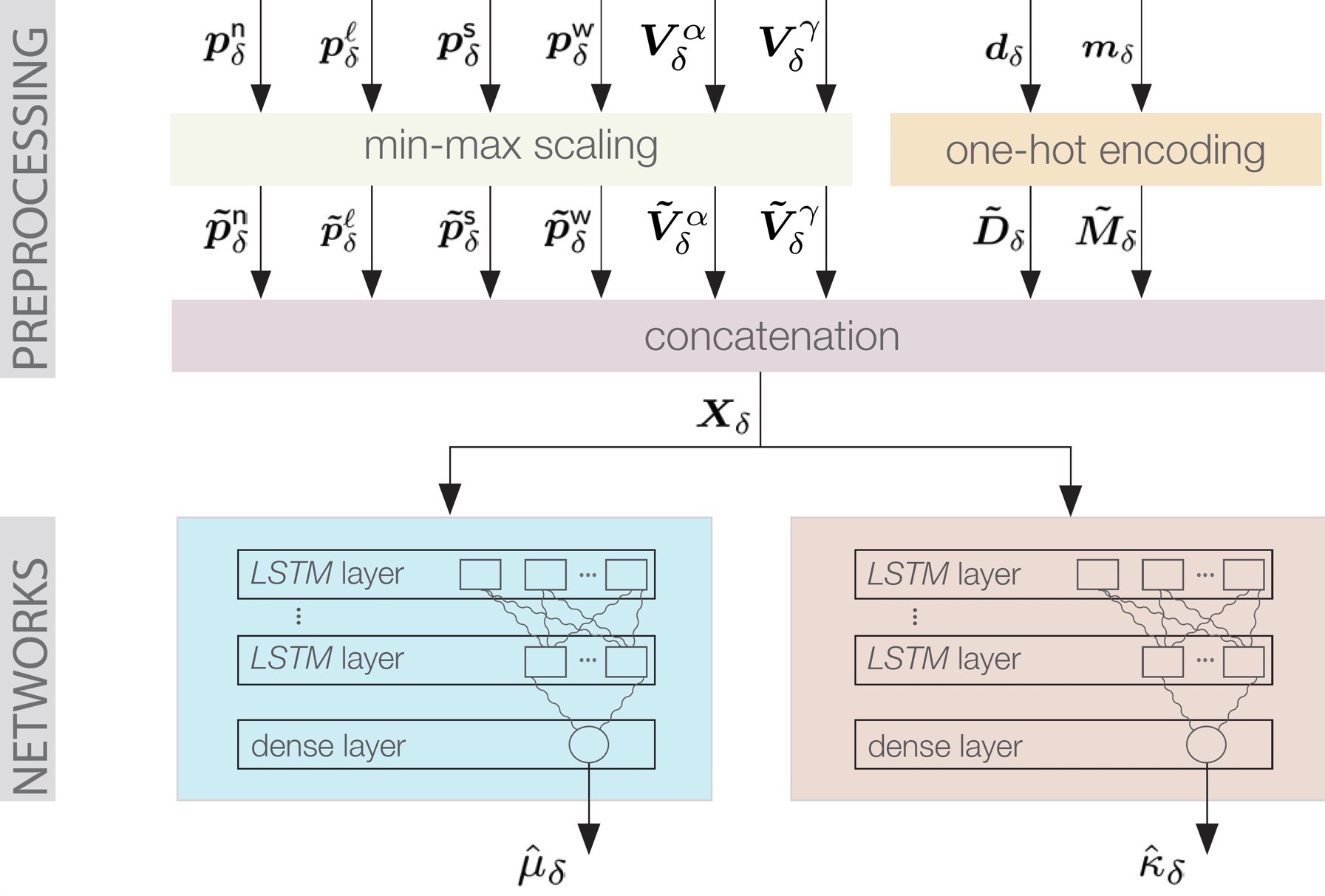}
\vspace{-0.4cm}
\caption{Graphical depiction of the proposed forecasting methodology}
\vspace{-0.2cm}
\label{fig6}
\end{figure}
We denote by $\boldsymbol{p}^{\mathsf{n}}_{\delta}$ the vector of historical net load values for $\mathcal{T}$ time periods, where the subscript $\delta$ is the index of the day for which the magnitude and start time of the primary three-hour net load ramp are forecast. We apply \textit{min-max scaling} to scale each element of $\boldsymbol{p}^{\mathsf{n}}_{\delta}$ to the range $[-1,1]$ and define by $\boldsymbol{\tilde{p}}^{\mathsf{n}}_{\delta}$ the vector of values obtained after the application of min-max scaling to each corresponding element of $\boldsymbol{p}^{\mathsf{n}}_{\delta}$. Let $\boldsymbol{p}^{\mathsf{\ell}}_{\delta}, \boldsymbol{p}^{\mathsf{s}}_{\delta} \text{, and } \boldsymbol{p}^{\mathsf{w}}_{\delta}$ denote the vectors of historical system load, \textit{PV} generation, and wind generation measurements for $\mathcal{T}$ time periods, respectively. We similarly apply min-max scaling to scale each element of $\boldsymbol{p}^{\mathsf{\ell}}_{\delta}, \boldsymbol{p}^{\mathsf{s}}_{\delta} \text{, and } \boldsymbol{p}^{\mathsf{w}}_{\delta} $ to the range $[-1,1]$ and denote by $\boldsymbol{\tilde{p}}^{\mathsf{\ell}}_{\delta}, \boldsymbol{\tilde{p}}^{\mathsf{s}}_{\delta} \text{, and } \boldsymbol{\tilde{p}}^{\mathsf{w}}_{\delta}  $ the vectors of values obtained after min-max scaling. Let $\boldsymbol{V}^{{\alpha}}_{\delta} \text{ and } \boldsymbol{V}^{{\gamma}}_{\delta}$ represent the matrices of historical temperature and solar irradiance measurements from the selected weather stations, respectively. We similarly apply min-max scaling to each corresponding element of $\boldsymbol{V}^{{\alpha}}_{\delta} \text{ and } \boldsymbol{V}^{{\gamma}}_{\delta} $ and obtain the matrices $\boldsymbol{\tilde{V}}^{{\alpha}}_{\delta} \text{ and } \boldsymbol{\tilde{V}}^{{\gamma}}_{\delta} $. We denote by $\boldsymbol{d_{\delta}}$ the day of the week and by $\boldsymbol{m_{\delta}}$ the month of the year of the selected $\mathcal{T}$ time periods. We express $\boldsymbol{d_{\delta}}$ and $\boldsymbol{m_{\delta}}$  in \textit{one-hot encoding} by the matrices $\boldsymbol{\tilde{D}}_{\delta} $ and $\boldsymbol{\tilde{M}}_{\delta}$, respectively.\par
We define by $\boldsymbol{X}_{\delta} \coloneqq \big[\boldsymbol{\tilde{p}}^{\mathsf{n}}_{\delta}, \boldsymbol{\tilde{p}}^{\mathsf{\ell}}_{\delta}, \boldsymbol{\tilde{p}}^{\mathsf{s}}_{\delta}, \boldsymbol{\tilde{p}}^{\mathsf{w}}_{\delta}, \boldsymbol{\tilde{V}}^{{\alpha}}_{\delta}, \boldsymbol{\tilde{V}}^{{\gamma}}_{\delta}, \boldsymbol{\tilde{D}}_{\delta}, \boldsymbol{\tilde{M}}_{\delta}\big]$
the input matrix of the two \textit{LSTM} networks. Each row of the matrix $\boldsymbol{X}_{\delta}$ is input sequentially in $\mathcal{T}$ steps to both \textit{LSTM} networks separately. To construct the networks, we utilize \textit{LSTM} blocks to form an \textit{LSTM} layer and---owing to the sequential nature of \textit{LSTM}s---stack one or multiple \textit{LSTM} layers to construct an \textit{LSTM} network. \par
For the \textit{LSTM} network to forecast the magnitude, we feed the output of the \textit{LSTM} blocks of the topmost layer to a conventional feedforward neural network with a single neuron, whose output is the forecast magnitude of the primary three-hour net load ramp on the subsequent day $\delta$, denoted by $\hat{\mu}_{\delta}$. Let $\mu_{\delta}$ denote the actual magnitude of the primary three-hour net load ramp on day $\delta$. We train the neural network to forecast the magnitude of the primary three-hour net load ramp with $\mathcal{N}$ training samples to minimize the loss function $\sum\limits_{{\delta}=1}^{\mathcal{N}} {\big({\mu_\delta}-\hat{\mu}_\delta\big)}^{2}$.\par
Analogously, for the \textit{LSTM} network to forecast the start time, we feed the output of the \textit{LSTM} blocks of the topmost layer to a conventional feedforward neural network with a single neuron. The proposed methodology relies on data/information reported using snapshots, where each snapshot represents certain conditions over a time period of specific duration. To ensure consistency with the time granularity with which data/information are reported, we round up the output of the conventional feedforward neural network to the nearest integer, which represents the time period at which the primary three-hour net load ramp on the subsequent day $\delta$ starts, denoted by $\hat{\kappa}_{\delta}$. Let $\kappa_{\delta}$ denote the actual time period at which the primary three-hour net load ramp on day $\delta$ starts. We train the neural network to forecast the start time of the primary three-hour net load ramp with $\mathcal{N}$ training samples to minimize the loss function $\sum\limits_{{\delta}=1}^{\mathcal{N}} {\big({\kappa_\delta}-\hat{\kappa}_\delta\big)}^{2}$.
\section{Case Study and Results}\label{4}
We illustrate the application of the proposed model to forecast the magnitude and start time of daily primary three-hour net load ramps in the \textit{CAISO} system.
\begin{figure*}[h]
\begin{minipage}[c][5.05cm][t]{\linewidth}
\includegraphics[width=18.2cm, height=5.0cm]{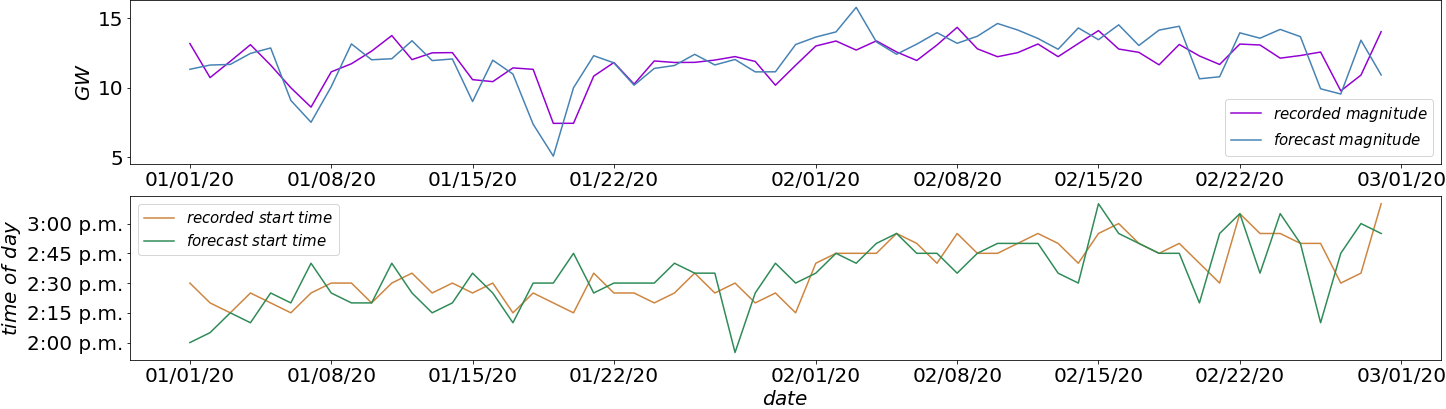}
\vspace{-0.65cm}
\caption{Forecast and recorded daily primary three-hour net load ramps}
\vspace{0cm}
\label{fig7}
\end{minipage}
\end{figure*}
\subsection{Dataset}
The original dataset \cite{rampbib:12} contains \textit{CAISO} historical net load, system load, \textit{PV} generation, and wind generation measurements at five-minute resolution, and we use the data collected from April 10, 2018 to February 29, 2020 to construct the dataset for our experiments. Further, we utilize the ambient temperature \cite{rampbib:14} and solar irradiance \cite{rampbib:13} measurements collected from the weather stations stated in Section \ref{2}. We sequentially split the dataset into training (70\%), validation (15\%), and test (15\%) sets.	
\subsection{Benchmarks}
Due to the scarcity of methods to forecast net load ramps, we utilize different \textit{RNN} and empirical models as benchmarks in the evaluation of the proposed model (\textit{PM}). A benchmark we use is the simple recurrent network (\textit{SRN}) model, which is a rudimentary \textit{RNN} architecture that lacks the gate structures of more sophisticated \textit{RNN} models. \par
We further utilize the \textit{GRU} model as a benchmark. The \textit{GRU} model is an \textit{RNN} architecture furnished with forget and reset gates. Despite its fewer internal structural elements, the \textit{GRU} model is widely touted as performing on par with the \textit{LSTM} model. As an empirical benchmark, we leverage the \textit{naïve persistence model} (\textit{NPM}), which forecasts that the magnitude and start time observed on any given day will also be observed on the subsequent day.
\subsection{Experimental Results}
We utilize Tensorflow and Keras to train, validate, and test the constructed neural networks.
All networks are trained with the \textit{Adam} optimizer and are provided with measurements collected from 12 p.m. to 8 p.m. of the previous day with 5-minute granularity, i.e., $\mathcal{T}=96$. We use the validation set to tune the hyperparameters of each network. \par
For the \textit{LSTM} network to forecast the magnitude of daily three-hour net load ramps, we decide on a learning rate of $3(10^{-3})$ and a three-layer architecture; the first layer contains 512 blocks and has a dropout rate of 0.15, the second layer contains 1024 blocks and has a dropout rate of 0.35, and the third layer contains 256 blocks and has a dropout rate of 0.40. Further, for the \textit{LSTM} network to forecast the start time of daily three-hour net load ramps, we settle on a three-layer architecture and a learning rate of $2.831(10^{-5})$; the first layer contains 128 blocks and has a dropout rate of 0.20, the second layer contains 256 blocks and has a dropout rate of 0.10, the third layer contains 32 blocks and has a dropout rate of 0.10. \par
We evaluate the performance of the models based on the mean squared error (\textit{MSE}), the mean absolute error (\textit{MAE}) and the mean absolute percentage error (\textit{MAPE}) metrics and tabulate the magnitude and start time forecasting results in Table \ref{magnperf} and Table \ref{stperf}, respectively. Each ``mean'' (resp. ``std'') row indicates the unweighted average (resp. standard deviation) of all forecast errors based on the corresponding metric.\par
\begin{table}[h]
\centering
\vspace{-0.2cm}
\footnotesize
{\fontsize{7.3}{8.76}\selectfont
\renewcommand{\arraystretch}{1.5}
\caption{Magnitude Forecasting Performances}
\label{magnperf}
\centering
\begin{tabular}{c | c || c | c | c | c }
\hline \hline
\multicolumn{2}{c ||}{metric} & \textit{PM} & \textit{GRU} & \textit{SRN} & \textit{NPM}\\
\hline \hline
\multirow{2}{*}{{\textit{MSE} }} &mean & 2.10095 & 2.84316 & 3.05268 & 3.36615\\
&std & 2.90149 &	4.28591	& 3.64901 & 	5.05616  \\
\hline 
\multirow{2}{*}{{\textit{MAE} }} &mean & 1.17132	&	1.31876	&	1.46620	&	1.39196 \\
&std & 1.43965	 &	1.65568	&	1.55531	&	1.83469 \\
\hline 
\multirow{2}{*}{{\textit{MAPE} }} &mean & 10.41129	&	11.87069	&	12.87822	&	13.00497\\
&std & 9.74688	&	12.05465&		11.36168	&	15.20778 \\
 \hline \hline
\end{tabular}}
\vspace{-0.1cm}
\end{table}
The reported values in Table \ref{magnperf} show that, in forecasting the magnitude of daily primary three-hour net load ramps, the proposed model yields lower mean errors than do the benchmarks based on all metrics. The standard deviation of the forecast errors obtained with the proposed model is also lower than that of the benchmarks, which bears out that the forecast errors of the proposed model are more clustered around the mean values and less spread out. On the heels of the proposed model is the \textit{GRU} model, which, in line with our expectations, outperforms the other benchmarks. We further note that the \textit{SRN} model results in higher mean values compared with the other \textit{RNN} architectures, which point to the shortcomings of the \textit{SRN} model to learn long-term temporal dependencies.\par
\begin{table}[h]
\vspace{-0.0cm}
\centering
\footnotesize
{\fontsize{7.3}{8.76}\selectfont
\renewcommand{\arraystretch}{1.5}
\caption{Start time Forecasting Performances}
\label{stperf}
\centering
\begin{tabular}{c | c || c | c | c | c }
\hline \hline
\multicolumn{2}{c ||}{metric} & \textit{PM} & \textit{GRU} & \textit{SRN} & \textit{NPM}\\
\hline \hline
\multirow{2}{*}{{\textit{MSE}}} &mean & 7.40572 &	9.51520	&22.82354	&10.93220\\
&std & 12.08975	& 13.98188 &	23.46784	& 16.80281 \\
\hline 
\multirow{2}{*}{{\textit{MAE} }} &mean & 2.07172 &	2.46524&	4.05900	&2.49153\\
&std & 2.71255	& 3.08281	& 4.26988	& 3.30113 \\
\hline 
\multirow{2}{*}{{\textit{MAPE} }} &mean & 1.16985	 & 1.39135	& 2.26724	& 1.40314\\
&std &1.00943	& 1.06318	& 1.39342&	1.22778 \\
 \hline \hline
\end{tabular}}
\vspace{-0.0cm}
\end{table}
In a similar vein, the reported mean and standard deviation values in Table \ref{stperf} certify that the proposed model outperforms the benchmarks in forecasting the start time of daily primary three-hour net load ramps based on all metrics. These results, in conjunction with the reported performances in Table \ref{magnperf}, reinforce our observation that the proposed methodology can successfully forecast daily primary three-hour net load ramps. We remark upon the lower error values of the \textit{GRU} model compared with the other benchmarks, which demonstrates the capability of the \textit{GRU} model to learn long-term temporal dependencies.\par
Fig. \ref{fig7} depicts the forecast magnitude and start time of daily primary three-hour net load ramps for a representative interval in the test set along with the actually recorded values. The plots in Fig. \ref{fig7} corroborate the numerical results reported in Table \ref{magnperf} and Table \ref{stperf} and visually verify that the proposed model, by and large, generates magnitude and start time forecasts that can effectively follow the actual values. We further note that both \textit{LSTM} networks of the proposed methodology require a computation time in the order of milliseconds to perform a single forecast on Intel\textsuperscript{\textregistered}\;Core\texttrademark\;i7-7500U \textit{CPU} @ 2.70 \textit{GHz} with 8 \textit{GB} of \textit{RAM}, which clearly permits the utilization of the proposed methodology in real-life applications. On these grounds, we maintain that the proposed methodology may lend itself as a powerful tool to forecast the magnitude and start time of daily primary three-hour net load ramps.
\enlargethispage{0cm}
\section{Conclusion}\label{5}
In this paper, we propose a methodology to forecast the magnitude and start time of the primary three-hour net load ramp on the subsequent day. The proposed methodology could prove useful for \textit{SO}s in meeting steep and sustained net load ramps. In particular, \textit{SO}s may capitalize on the forecasts to evaluate the required flexible capacity so as to maintain the supply-demand balance around the clock. We demonstrate the application of the proposed methodology on the \textit{CAISO} system. The results illustrate that the proposed methodology generates forecasts that follow the general trend of net load ramps and outperforms the selected benchmarks based on various evaluation metrics. \par
A natural extension of the presented work is to forecast the magnitude and start time of daily primary three-hour net load ramps with shorter lead times by utilizing the measurements collected on the day for which the forecasts are performed.

\end{document}